\documentclass{article}
\usepackage{spconf,amsmath,graphicx,enumitem,setspace,tabularx,booktabs,multirow}
\usepackage[font=footnotesize]{subfig}
\setitemize[1]{topsep=0pt,itemsep=0.5pt,parsep=0pt,leftmargin=2em}
\usepackage{color}
\definecolor{gray}{rgb}{0.5,0.6,0.7}
\title{Audio-Visual Quality Assessment for User Generated Content: \\Database and Method}
\name{Yuqin Cao$^{\star}$ \qquad Xiongkuo Min$^{\star}$ \qquad Wei Sun$^{\star}$ \qquad Xiaoping Zhang$^{\dag}$ \qquad Guangtao Zhai$^{\star}$
\thanks{Corresponding authors: Xiongkuo~Min and Guangtao~Zhai.}
\thanks{Thanks for the support provided by OpenI Community (https://openi.pcl.ac.cn).}}
\address{$^{\star}$Institute of Image Communication and Information Processing, Shanghai Jiao Tong University\\
$^{\dag}$ Department of Electrical, Computer and Biomedical Engineering, Toronto Metropolitan
University}

\begin{document}

\maketitle

\begin{abstract}
With the explosive increase of User Generated Content (UGC), UGC video quality assessment (VQA) becomes more and more important for improving users’ Quality of Experience (QoE). However, most existing UGC VQA studies only focus on the visual distortions of videos, ignoring that the user's QoE also depends on the accompanying audio signals. In this paper, we conduct the first study to address the problem of UGC audio and video quality assessment (AVQA). Specifically, we construct the first \textbf{U}GC \textbf{AV}QA database named the SJTU-UAV database, which includes 520 in-the-wild UGC audio and video (A/V) sequences, and conduct a user study to obtain the mean opinion scores of the A/V sequences. The content of the SJTU-UAV database is then analyzed from both the audio and video aspects to show the database characteristics. We also design a family of AVQA models, which fuse the popular VQA methods and audio features via support vector regressor (SVR). We validate the effectiveness of the proposed models on the three databases. The experimental results show that with the help of audio signals, the VQA models can evaluate the perceptual quality more accurately. The database will be released to facilitate further research.
\end{abstract}
\begin{keywords}
User-generated content, audio-visual quality assessment, multimodal fusion
\end{keywords}
\section{Introduction}
More and more users capture and share their own audio and video (A/V) on various social media platforms. This type of A/V is known as user-generated content (UGC) A/V. Since UGC A/V are typically shot by inexperienced users, it always suffers from a variety of authentic distortions, such as video distortions (like noise, camera shake, and over/under-exposure) and audio distortions (like background noise, wind noise, and handling noise). These distortions may degrade the user's Quality of Experience (QoE). Though many attentions have been paid to UGC VQA problems \cite{sun2021deep,min2021screen,Sunugc}, there is no study to investigate the problem of UGC AVQA, which is more practical in real-world scenarios.

In the past few decades, many AVQA databases have been proposed. Most AVQA databases are created by artificially degrading a few high-quality A/V sequences with a variety of different A/V distortion types. These kinds of A/V databases are referred to as synthetically-distorted AVQA databases. Compared with UGC AVQA databases, they are too simple to represent the authentic distortions existing in real-world A/V sequences. As far as we know, the UGC AVQA database has not yet been presented and there is an urgent need to construct a UGC AVQA database.

In this paper, we make three contributions to the UGC AVQA field. Firstly, we construct the first public UGC AVQA database, called the SJTU-UAV database, which contains 520 UGC A/V sequences. We designed and conducted a subjective experiment to assess the quality of each A/V sequence in the SJTU-UAV database. Secondly, we characterize the content diversity of the SJTU-UAV database in terms of five video attributes and four audio attributes and compare it with two synthetically-distorted AVQA datasets, i.e. the LIVE-SJTU database \cite{min2020study} and the UnB-AVC database \cite{martinez2014no}. Thirdly, we design a family of UGC AVQA models by fusing existing UGC VQA models and representative audio features via support vector regressor (SVR). We test and compare these models on the SJTU-UAV database, the LIVE-SJTU database, and the UnB-AVC database. The experimental results show that most UGC VQA models with audio features have better performances than that without audio features.

\begin{table*}[!t]
    \vspace{-0.5cm}
    \centering
    \caption{The basic information of three databases: SJTU-UAV, LIVE-SJTU, and UnB-AVC.}
    \resizebox{0.8\textwidth}{!}{
    \begin{tabular}{c| cccccccc}
    \toprule
    Database & Unique A/V contents & Total A/V & Distortion Type & Resolution & Length \\
    \midrule
    SJTU-UAV & 520 & 520 & In-the-wild & $960\times 720$ - $1920\times1080$ & 8s\\
    LIVE-SJTU \cite{min2020study}& 14 & 336 & Video compression, audio compression& $1920\times1080$ & 8s \\
    UnB-AVC \cite{martinez2014no}& 6 & 72 & Video compression, audio compression& $1280\times720$ & 8s \\
    \bottomrule
    \end{tabular}
    }
    \label{tab:Database}
\end{table*}
\vspace{-0.2cm}
\section{SJTU-UAV Database}
To facilitate the study on UGC AVQA, we collected 520 UGC A/V sequences from the YFCC100m database \cite{thomee2016yfcc100m} to construct the SJTU-UAV database, the first UGC AVQA database. A subjective experiment was conducted to obtain the mean opinion scores (MOSs) of the SJTU-UAV database.
\vspace{-0.2cm}
\subsection{UGC A/V Sequences Collection}
To collect UGC A/V sequences with diverse scenes, we first searched 65 key terms on the YFCC100m database \cite{thomee2016yfcc100m}. Then, we chose 520 A/V sequences based on some practical considerations, such that they: 1) were still available for download, 2) had an audio channel, 3) lasted longer than 8 seconds, 4) were in landscape layout for further processing. We cut 8 seconds clips from A/V sequences and resized the resolution of the video channel until the resolution of the video channel was no smaller than $960 \times 720$. The magnitude of audio channels was normalized to avoid frequent volume adjustment during the test and made human subjects concentrate on the subjective experiments.

After the above steps, the SJTU-UAV database consisted of 520 UGC A/V sequences, with the largest percentage of videos having a resolution of $1280\times720$ pixels (45\% of the videos), followed by $960\times720$ (42\%) and $1920\times1080$ (8\%). The frame rates of the video channels vary from $15$ to $30$ frames per second. The sampling rates of the audio channels range from $44.1$ kHz to $96$ kHz. 
\subsection{Subjective Quality Assessment Study}
We conducted a subjective experiment to obtain the MOSs of the SJTU-UAV database. The A/V testing environment included a room equipped with a Redmi 23.8-inch $1920\times1080$ monitor and a Sony WH-1000XM4 headphone.

The single stimulus continuous quality evaluation (SSCQE) strategy was adopted in the subjective experiments. A total of 21 college students participated in the experiment. Every subject first read an instruction explaining the experimental process and experiment requirements, and then experienced a brief training in which subjects watched 10 A/V sequences (not included in the formal test) to become familiar with the operation procedure and the distortions that may occur. Then the formal experiment was accomplished by two sessions, where each session contained 260 A/V sequences. The order of the test videos was random for each subject to avoid bias. All A/V sequences were displayed at native resolution. After each A/V sequence was viewed, a continuous quality rating bar on the user interface was presented to the subject, which would be labeled with five Likert adjectives: Bad, Poor, Fair, Good, and Excellent. The subject dragged a slider along the continuous quality bar to give an opinion score of the overall A/V quality they perceived.
\begin{figure}[t]
 \centering
    \includegraphics[width=.3\textwidth]{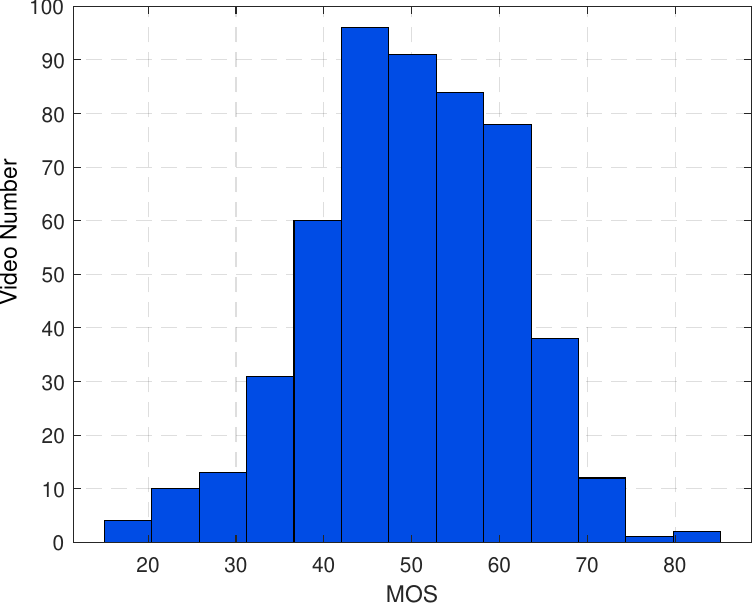}
\caption{Histogram of MOSs from the SJTU-UAV database.}
\label{fig:MOS}
\vspace{-0.5cm}
\end{figure}
\subsection{Subjective Data Processing}
We followed the subjective data processing method recommended by \cite{bt2002methodology}. None of the 21 subjects was identified as an outlier and eliminated. We normalized the raw scores of subjects to Z-scores ranging between 0 and 100 and calculated the mean of Z-scores to obtain the MOSs. Fig. \ref{fig:MOS} illustrates the histogram of MOSs over the entire database, showing a broad MOS distribution of A/V sequences. 

\begin{figure*}[!tb]
\vspace{-0.7cm}
\captionsetup[subfigure]{justification=centering}
\centering
   \subfloat[Contrast]{\includegraphics[width=.15\linewidth]{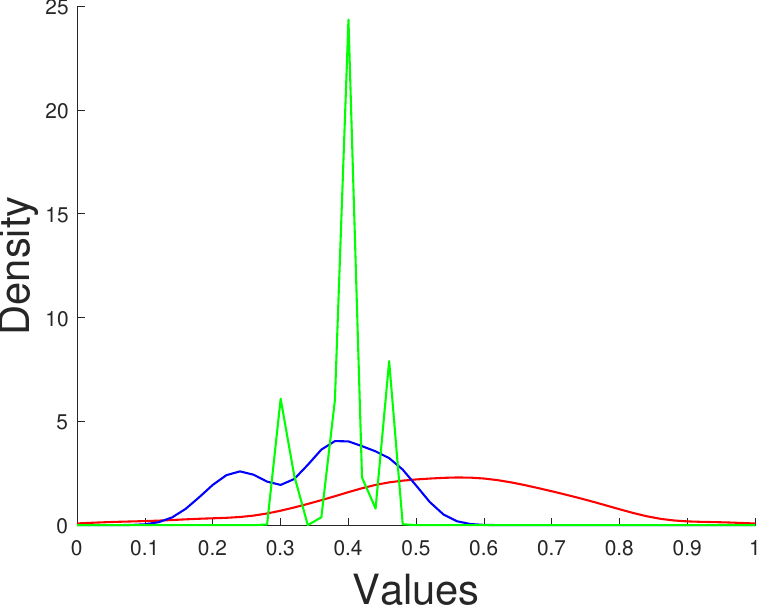}
\label{fig:contrast}}
   \subfloat[Colorfulness]{\includegraphics[width=.15\linewidth]{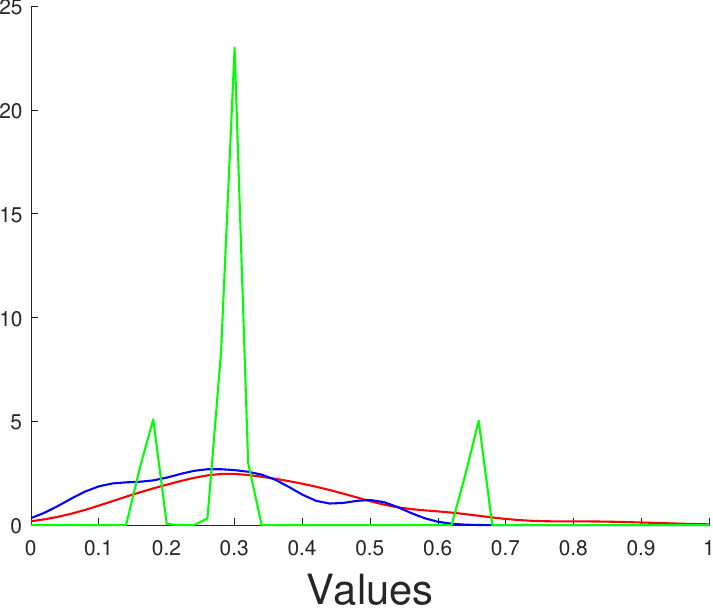}
\label{fig:color}}
   \subfloat[CPBD]{\includegraphics[width=.15\linewidth]{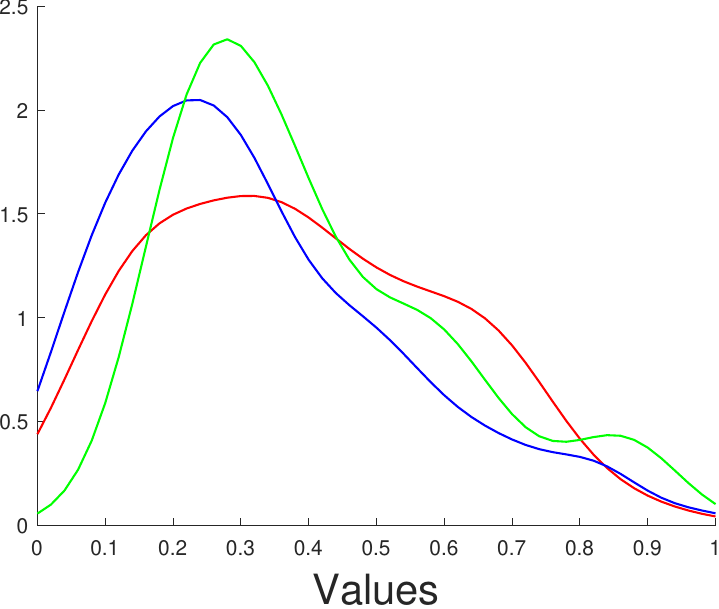}
\label{fig:CPBD}}
   \subfloat[SI]{\includegraphics[width=.15\linewidth]{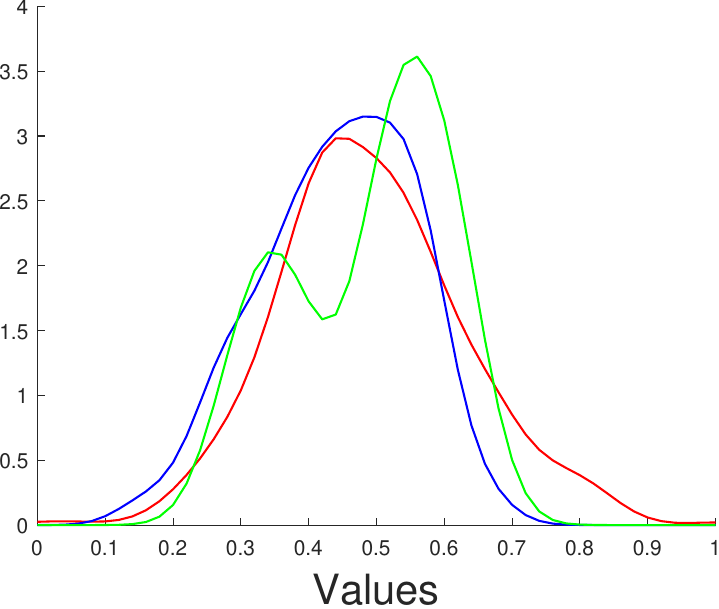}
\label{fig:SI}}
   \subfloat[TI]{\includegraphics[width=.15\linewidth]{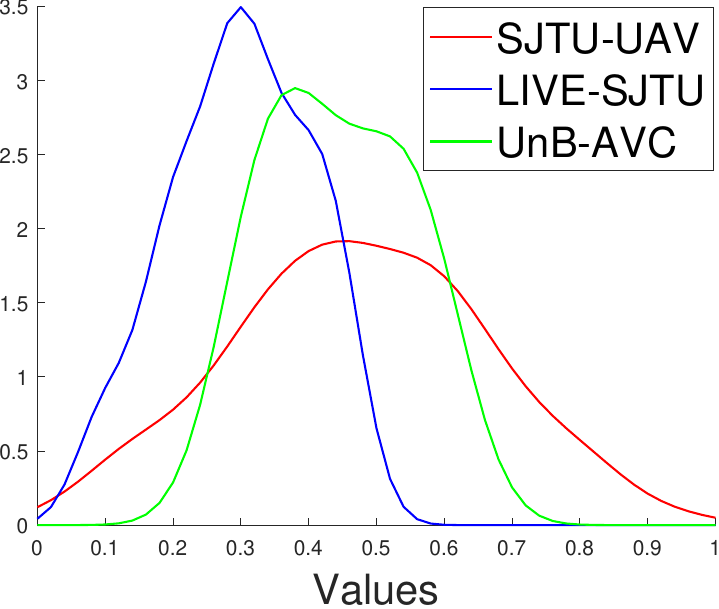}
\label{fig:TI}}
\\
\vspace{-0.2cm}
   \subfloat[SEF]{\includegraphics[width=.15\linewidth]{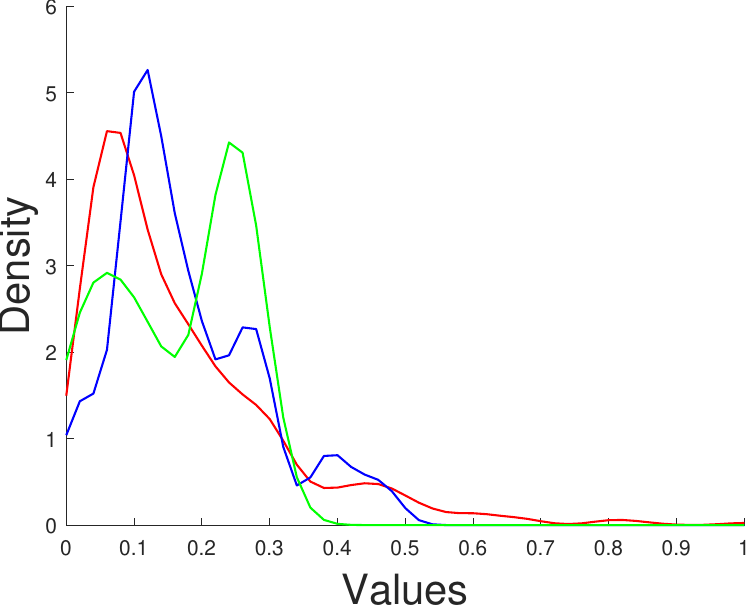}
\label{fig:SEF}}
   \subfloat[ZCR]{\includegraphics[width=.15\linewidth]{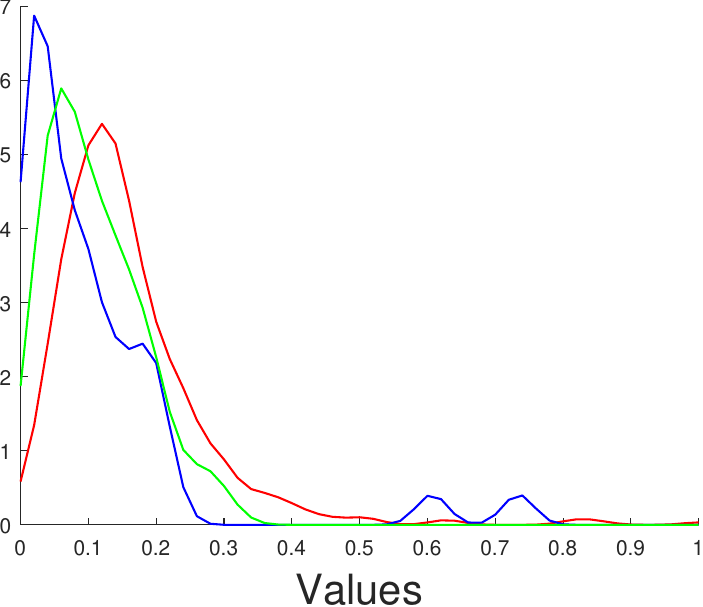}
\label{fig:ZCR}}
   \subfloat[SC]{\includegraphics[width=.15\linewidth]{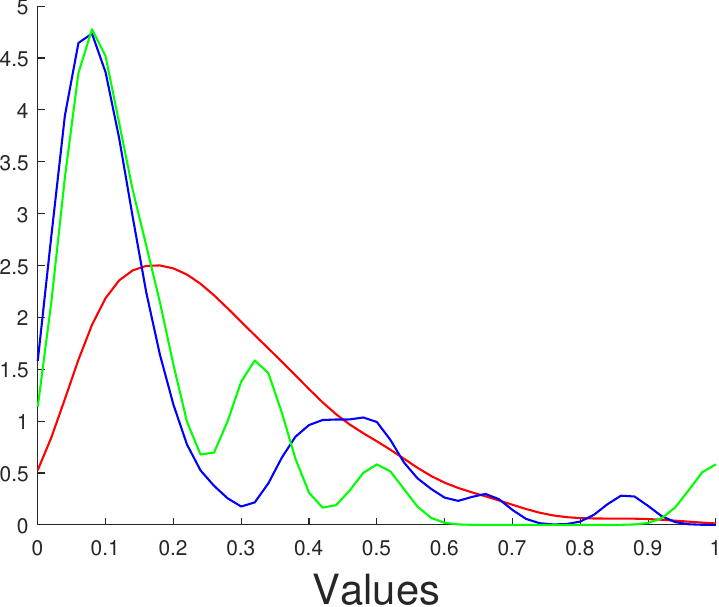}
\label{fig:SC}}
   \subfloat[SE]{\includegraphics[width=.15\linewidth]{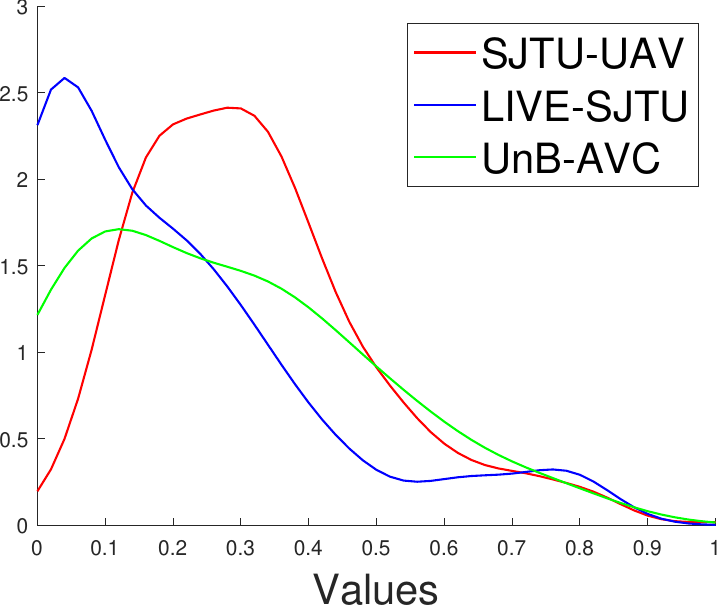}
\label{fig:SE}}
   \caption{Distribution comparisons of A/V attributes on the three databases: SJTU-UAV, LIVE-SJTU, and UnB-AVC.}
\label{fig:video attribute}
% \vspace{-0.3cm}
\end{figure*}
\section{AVQA Database Attribute Analysis}
In order to characterize the content diversity of the A/V sequences in the SJTU-UAV database, we analyze five video attributes and four audio attributes. Since the SJTU-UAV database is the first UGC AVQA database, we select two synthetically-distorted AVQA databases, the LIVE-SJTU database \cite{min2020study}, and the UnB-AVC database \cite{martinez2014no}, for comparison. Table \ref{tab:Database} summarizes the main information of three databases. The LIVE-SJTU database includes 336 A/V sequences, which are generated from 14 original source contents by applying 24 different A/V distortion combinations on them. The UnB-AVC database includes 72 A/V sequences that are generated by applying 12 different A/V distortion combinations on 6 original source contents.
\vspace{-0.1cm}
\subsection{Video Attributes}
\vspace{-0.1cm}
We select contrast, colorfulness, cumulative probability of blur detection (CPBD), spatial information (SI), and temporal information (TI) as video attributes. All video attributes are calculated on every $10$-th frame to reduce computational complexity and then averaged over frames of each A/V sequence.
\begin{itemize} 
    \itshape \item Contrast: \upshape The contrast metric is measured simply by the std of pixel gray-scale intensities \cite{hasler2003measuring}.
    \itshape \item Colorfulness: \upshape The colorfulness metric is reported in \cite{hasler2003measuring}. With the RGB channels of a frame as matrices $R$, $G$, and $B$, we first compute two matrices $rg=R-G$ and $yb=\frac{1}{2}(R+G)-B$. Then, the metric is calculated as $\sqrt{\sigma_{rg}^2+\sigma_{yb}^2}+\sqrt{\mu_{rg}^2+\mu_{yb}^2}$, where $\sigma$ and $\mu$ denote the std and mean of their respective matrices respectively.
    \itshape \item CPBD: \upshape The CPBD metric \cite{narvekar2011no} is a perceptual NR image sharpness metric, which can estimate the probability of detecting blur at each edge in the frame.
    \itshape \item SI: \upshape SI is obtained by applying a Sobel filter to each frame to extract the gradient magnitude $Gmag$ and the gradient direction $Gdir$ for each pixel and then the metric is calculated as $\frac{1}{2}(\sigma_{Gmag}+\sigma_{Gdir})$, where $\sigma_{Gmag}$ and $\sigma_{Gdir}$ denote the std of $Gmag$ and $Gdir$.
    \itshape \item TI: \upshape TI calculates stds of pixel-wise frame difference.
\end{itemize}
\vspace{-0.2cm}
\subsection{Audio Attributes}
For audio attributes, we choose short-term energy fluctuation (SEF), zero-crossing rate (ZCR), spectral centroid (SC), and spectral entropy (SE), which have been widely used in various audio analysis approaches \cite{giannakopoulos2014introduction}. SEF and ZCR analyze audio signals in the time domain, while SC and SE are computed in the frequency domain.

\begin{itemize}
    \itshape \item SEF: \upshape SEF describes the variation of successive frames. A higher SEF means audio signals alternate rapidly between high and low energy states such as speech signals.
    \itshape \item ZCR: \upshape ZCR is the frequency of times signals' values change from positive to negative and vice versa, which can measure the noisiness of audio signals.
    \itshape \item SC: \upshape A higher value of SC \cite{giannakopoulos2014introduction} corresponds to brighter sounds. The maximum value of SC for all audio frames yields SC of the A/V sequence.
    \itshape \item SE: \upshape The std of SE for all audio frames yields the SE of the whole audio. The environmental sounds always have lower SE values, while speech segments yield the higher SE values.
\end{itemize}
\vspace{-0.2cm}
\subsection{Observations}
The distributions of video attributes and audio attributes on each database are shown in Fig. \ref{fig:video attribute}, respectively. In term of contrast and colorfulness, the SJTU-UAV database and the LIVE-SJTU database are more evenly distributed, while the UnB-AVC database is centered on a few peaks, because the UnB-AVC database only has 6 original sources. For the distributions of CPBD, SI, and TI, the SJTU-UAV database spreads most widely among the three databases. For audio attributes, all three databases adhere closer to small values. While for the SC distribution, the SJTU-UAV database spreads most widely, indicating diverse tone qualities of audio signals, which is consistent with the observations that the SJTU-UAV database has more various types of audio signals, such as music, wind, human voice, crowd, and traffic sound, etc.
On the whole, the SJTU-UAV database is more content-diverse, uniformly-distributed than the LIVE-SJTU database and the UnB-AVC database.

\begin{table*}[!t]
\vspace{-0.5cm}
    \centering
    \caption{Mean SRCC and PLCC of the UGC AVQA models on the three databases. The best and top $3$ performances for each database are marked in boldface and underlined, respectively. ``W/O AF'' indicates without audio features. ``RP'' indicates RASTA-PLP.}
    \label{tab:all performance}
    \vspace{-0.2cm}
    \resizebox{\textwidth}{!}{
    \begin{tabular}{cc| ccccc| ccccc| ccccc}
    \toprule 
    \multirow{2}{*}{Criteria} & Video & \multicolumn{5}{c|}{SJTU-UAV} & \multicolumn{5}{c|}{LIVE-SJTU}& \multicolumn{5}{c}{UnB-AVC}\\ ~ & Model & W/O AF & MFCC & RP & NRMusic & ARN-50 & W/O AF & MFCC & RP & NRMusic & ARN-50 & W/O AF & MFCC & RP & NRMusic & ARN-50\\
    \midrule
    \multirow{7}{*}{SRCC} & 
    BRISQUE & 
    0.6664 & 0.6570 & 0.6351 & 0.6597 & 0.6611 & 
    0.5959 & 0.7201 & 0.8548 & 0.8383 & 0.8501 & 
    0.5614 & 0.6143 & 0.6274 & 0.7237 & 0.7192\\
    ~&VRN-50 & 
    0.7386 & 0.7401 & 0.7394 & \underline{0.7412} & \underline{\textbf{0.7492}} & 
    0.7049 & 0.7655 & \underline{0.9302} & 0.9108 & \underline{\textbf{0.9317}} &
    0.8089 & 0.8431 & \underline{0.9038} & \underline{\textbf{0.9134}} & \underline{0.8878}\\
    ~&CORNIA & 
    0.7304 & \underline{0.7412} & 0.7295 & \underline{0.7481} & 0.7387 & 
    0.7238 & 0.8645 & 0.9235 & 0.9089 & \underline{0.9251} & 
    0.7521 & 0.8520 & 0.8054 & 0.8485 & 0.8380 \\
    ~&V-BLIINDS & 
    0.7160 & 0.7177 & 0.7041 & 0.7160 & 0.7178 & 0.6681 & 0.7634 & 0.8651 & 0.8648 & 0.8833 & 0.5506 & 0.7697 & 0.7704 & 0.8323 & 0.8271\\
    ~&TLVQM & 
    0.5999 & 0.6185 & 0.6233 & 0.6163 & 0.6123 & 0.6103 & 0.6856 & 0.8275 & 0.7888 & 0.7962 & 0.7701 & 0.8793 & 0.8320 & 0.8624 & 0.8600\\
    ~&VIDEVAL & 
    0.7085 & 0.7225 & 0.7152 & 0.7198 & 0.7247 & 
    0.6598 & 0.7154 & 0.8608 & 0.8575 & 0.8453 & 0.6154 & 0.8716 & 0.8039 & 0.8744 & 0.8823\\
    ~&RAPIQUE & 
    0.7171 & 0.7265 & 0.7129 & 0.7303 & 0.7228 & 0.6219 & 0.7119 & 0.8320 & 0.8304 & 0.8112 & 0.8421 & 0.8543 & 0.8625 & 0.8800 & 0.8613\\
    
    \midrule
    \multirow{7}{*}{PLCC} & 
    BRISQUE & 
    0.6774 & 0.6664 & 0.6446 & 0.6621 & 0.6644 & 0.6203 & 0.7289 & 0.8618 & 0.8355 & 0.8547 & 0.7579 & 0.7427 & 0.7017 & 0.7775 & 0.7792\\
    ~&VRN-50 & 
    0.7540 & \underline{0.7554} & 0.7489 & 0.7522 & \underline{\textbf{0.7579}} & 
    0.7027 & 0.7683 & \underline{0.9387} & 0.9217 & \underline{\textbf{0.9423}} &
    0.8665 & 0.8606 & \underline{0.9073} & \underline{\textbf{0.9286}} & 0.8899\\
    ~&CORNIA & 
    0.7375 & 0.7493 & 0.7375 & \underline{0.7563} & 0.7453 &
    0.7218 & 0.7114 & 0.9319 & 0.9156 & \underline{0.9325} & 
    0.7962 & 0.8672 & 0.8419 & 0.8854 & 0.8469\\
    ~&V-BLIINDS & 
    0.7364 & 0.7356 & 0.7213 & 0.7285 & 0.7305 & 
    0.6804 & 0.7132 & 0.8744 & 0.8722 & 0.8877 & 0.7478 & 0.7988 & 0.8028 & 0.8202 & 0.8304\\
    ~&TLVQM & 
    0.6060 & 0.6218 & 0.6215 & 0.6133 & 0.6121 & 0.6247 & 0.8726 & 0.8267 & 0.7816 & 0.7878 & 0.8391 & 0.8613 & 0.8223 & 0.8370 & 0.8674 \\
    ~&VIDEVAL & 
    0.7256 & 0.7342 & 0.7279 & 0.7277 & 0.7279 & 0.6613 & 0.8004 & 0.8622 & 0.8596 & 0.8441 & 0.8387 & 0.8900 & 0.8411 & 0.8869 & 0.8963\\
    ~&RAPIQUE & 
    0.7304 & 0.7388 & 0.7235 & 0.7390 & 0.7325 & 0.6377 & 0.7326 & 0.8509 & 0.8338 & 0.8197 & 0.8442 & 0.8785 & 0.8594 & \underline{0.9183} & 0.8659\\
    \bottomrule
    \end{tabular}
    }
    %\bigskip
\end{table*}
\section{OBJECTIVE UGC AVQA}
\label{sec:UGC-AVQA Models}
Considering many single-mode NR I/VQA and AQA methods have been proposed, we design a family of UGC AVQA models. We first utilize several well-known UGC VQA models and several popular audio analysis methods to extract the corresponding visual and audio features. Then we fuse visual and audio features into the final scores via SVR \cite{chang2011libsvm}. Considering resolutions also influence the users' QoE, video resolutions are also utilized to train SVR. The models can be defined as:
\begin{equation}
    Q_{av} = SVR(w, h, f_v, f_a),
\end{equation}
where $f_v$ and $f_a$ denote video and audio quality-aware feature vectors, $w$ and $h$ denote the width and height of videos, $Q_{av}$ is the final score. We utilize the following handcrafted video and audio feature extractors:
\begin{itemize}
    \itshape \item Video: \upshape BRISQUE \cite{6272356}, CORNIA \cite{ye2012unsupervised}, V-BLIINDS \cite{saad2014blind}, TLVQM \cite{korhonen2019two}, VIDEVAL \cite{tu2021ugc}, and RAPIQUE \cite{tu2021rapique}.
    \itshape \item Audio: \upshape Mel frequency cepstral coefficient (MFCC), RASTA-PLP \cite{hermansky1991rasta} and NRMusic \cite{li2013non}.
\end{itemize}
The AQA models mentioned above extract features from each audio segment and calculate the means and stds over all audio segments to produce audio quality-aware features. The audio segments correspond with the closest video frames.

Given the excellent feature extraction ability of DNNs, we also employ DNNs as video and audio feature extractors. We design two models, named VRN-50 and ARN-50, which remove the last fully connected layer of ResNet-50 to extract video and audio features respectively, as recommended by \cite{min2020study,cao2021deep}. For VRN-50, we randomly crop $N$ patches from each video frame and then fed them into ResNet-50 to extract patch features. The patch features of all $N$ patches are then averaged to produce the video frame features, and then the mean of video frame features denotes the video quality-aware features. For ARN-50, we first utilize the short-time Fourier transform (STFT) to calculate the spectrogram of each audio segment. The spectrograms are fed into ResNet-50 to extract spectrogram features, which are then averaged over all audio segments to produce the audio features.

\section{Experiments}
We test the proposed UGC AVQA models introduced in Section \ref{sec:UGC-AVQA Models} on the three databases.
A total of $7$ (video models) $\times$ $4$ (audio models) $= 28$ models are tested and compared.

\subsection{Experimental Settings}
We conduct experiments on the authentically-distorted AVQA database and synthetically-distorted AVQA databases: SJTU-UAV, LIVE-SJTU, and UnB-AVC. We randomly split each database into a training set ($80\%$ of the A/V sequences) and a testing set ($20\%$ of the A/V sequences) with no overlap of video contents. For the LIVE-SJTU database and the UnB-AVC database, all distorted A/V sequences derived from the same reference A/V sequences are divided into the same set. We train models only on the training set and find the top models generating the lowest RMSE on the training set, and then test the top model on the testing set as the final model performance. All methods are trained and tested with the same training/testing splits. This procedure is repeated 100 times to prevent performance bias and the mean performance is recorded.
We use Spearman’s rank-order correlation coefficient (SRCC) and Pearson’s linear correlation coefficient (PLCC) to evaluate the effectiveness of AVQA methods. 

\subsection{Evaluation of UGC AVQA Models}
We compare the performances on the SJTU-UAV database, the LIVE-SJTU database, and the UnB-AVC database in Table \ref{tab:all performance}, from which we have several interesting observations. Firstly, most UGC VQA models with audio features achieve better performances than that without audio features. Secondly, the models combining the video models VRN-50, CORNIA, and VIDEVAL and the audio models NRMusic and ARN-50 yield relatively better performances, which suggests that VQA models and representative audio features can extract the quality-aware features to some extent for both authentically-distorted A/V sequences and synthetically-distorted A/V sequences. Thirdly, the model fusing the features of VRN-50 and ARN-50 shows the best performance on the SJTU-UAV database and LIVE-SJTU database, while having relatively poor performance on the UnB-AVC database. Differing from handcraft AQA models (MFCC, RASTA-PLP, and NRMusic), ARN-50 utilizes DNNs to extract audio features. Since the number of A/V sequences in the UnB-AVC database is much smaller, it may be more difficult for ARN-50 to predict quality scores. Finally, the audio models of the family of AVQA models for UGC A/V provide greater performance improvements on the LIVE-SJTU database and UnB-AVC database than on the SJTU-UAV database. It indicates that the influence of audio signals in the SJTU-UAV database is more difficult to learn.
\vspace{-0.2cm}
\section{CONCLUSIONS}
In this paper, we construct the first UGC AVQA database: the SJTU-UAV Database, which is composed of 520 UGC A/V sequences. A subjective experiment is conducted on the SJTU-UAV database to obtain the MOSs. Then, four audio and five video attributes are utilized to characterize the content diversity of the AVQA databases, which demonstrates that the SJTU-UAV database is more content-diverse, uniformly-distributed than the LIVE-SJTU database and the UnB-AVC database. Finally, we design a family of UGC AVQA models and test them on the three databases.
\bibliographystyle{main}
\bibliography{./main}

\end{document}